\documentclass[%
preprint,
 amsmath,amssymb,
 aps,
]{revtex4-2}

\usepackage{graphicx}
\usepackage{dcolumn}
\usepackage{bm}
\usepackage[version=3]{mhchem}
\usepackage{gensymb}
\usepackage{float}
\NewDocumentCommand{\MakeTitleInner}{ +m +m +m }{
    \newpage%
    \null%
    \vskip 2em%
    \begin{center}%
        \let \footnote \thanks
        {\LARGE #1 \par}
        \vskip 1.5em%
        {%
            \large
            \lineskip .5em%
            \begin{tabular}[t]{c}%
                #2
            \end{tabular}\par%
        }%
        \vskip 1em%
        {\large #3}
    \end{center}%
    \par
    \vskip 1.5em%
}
\NewDocumentCommand{\MakeTitle}{ +m +m +m }{%
    \begingroup
        \renewcommand\thefootnote{\@fnsymbol\c@footnote}%
        \def\@makefnmark{\rlap{\@textsuperscript{\normalfont\@thefnmark}}}%
        \long\def\@makefntext##1{\parindent 1em\noindent
            \hb@xt@1.8em{%
                \hss\@textsuperscript{\normalfont\@thefnmark}%
            }##1%
        }%
        \if@twocolumn
            \ifnum \col@number=\@ne
                \MakeTitleInner{#1}{#2}{#3}
            \else
                \twocolumn[\MakeTitleInner{#1}{#2}{#3}]%
            \fi
        \else
            \newpage
            \MakeTitleInner{#1}{#2}{#3}
        \fi
        \thispagestyle{plain}
    \endgroup
    \setcounter{footnote}{0}%
}
\makeatother

\begin{document}

\preprint{}

\title{Computer Simulation of the Growth of a Metal-Organic Framework Proto-crystal at Constant Chemical Potential} 

\author{Sahar Andarzi Gargari}
\author{Emilio Méndez}
\author{Rocio Semino}%
 \email{rocio.semino@sorbonne-universite.fr}
\affiliation{Sorbonne Université, CNRS, Physico-chimie des Electrolytes et Nanosystèmes Interfaciaux, PHENIX, F-75005 Paris, France.}%

\date{\today}

\begin{abstract}

Designing metal-organic frameworks (MOFs) synthesis protocols is currently largely driven by trial-and-error, since we lack fundamental understanding of the molecular level mechanisms that underlie their self-assembly processes. Previous works have studied the nucleation of MOFs, but  their growth has never been studied by means of computer simulations, which provide molecular level detail. In this work, we combine constant chemical potential simulations with a particle insertion method to model the growth of the ZIF-8 MOF at varying synthesis temperatures and concentrations of the reactants. 
Non-classical growth mechanisms triggered by oligomer attachments were detected, with a higher predominance in the most concentrated setups.
The newly formed layers preserve the pore-like density profile of the seed crystal but contain defective sites characterized by the presence of 3, 5 and 7 membered rings, typical of amorphous phases.
Compared to the amorphous intermediate species obtained at the nucleation part of the self-assembly process previously investigated in our group [Chem. Mater., doi: 10.1021/acs.chemmater.5c02028, 2025], larger-sized rings are more common in the grown layer. Moreover, these are favored by increasing reactant concentration and temperature, as is the degree of deviation with respect to the original crystal structure. We computed growth rates for the steady-state regime, and the non-linear tendency with respect to concentration leads us to hypothesize that in these conditions the growth is controlled by the adsorption rather than by the diffusion processes.

\end{abstract}

\maketitle


\section{Introduction}

Metal-organic frameworks (MOFs) have attracted the attention of materials scientists and entrepreneurs alike, due to their potential to revolutionize food conservation, pharmacological and depollution industries, among others.\cite{Khezerlou_2025,Nguyen_2024,Xu_2024} These crystalline solids with porosities that feature precisely engineered sizes and chemical functions ally confinement and chemical effects for diverse applications. They are formed by the self-assembly of metal ions or metal-containing oxoclusters with organic molecules via coordination bonds.

Understanding the self-assembly mechanisms of MOFs is crucial for unlocking their rational design, but this is a very challenging task, since self-assembly is a collection of chemical processes associated to a wide range of free energy barriers, and it relies on a delicate equilibrium between kinetic and thermodynamic factors.\cite{Cheetham2018} Some of these chemical processes are rare events and are thus difficult to sample. The chemical species that are formed along the process (building units, a series of polydisperse oligomers, growing and critical nuclei, intermediate phases and the final crystal) also feature a wide range of characteristic length-scales. This makes it impossible to probe the whole self-assembly process via a single experimental (or computational) technique. 

In the quest of better understanding MOF self-assembly, ZIF-8,\cite{Park2006} a MOF formed by Zn$^{2+}$ ions and 2-methylimidazolate (MIm$^-$) ligands, was given particular attention. Indeed, its facile synthesis, combined with its promising applications, have made this MOF an archetypal example for many fundamental science studies. ZIF-8 is a zeolitic-imidazolate framework (ZIF) which implies that it acquires a zeolite-like topology, with Zn-MIm-Zn angles around 145\degree. 

In the past decade, there were considerable advances in the direction of better understanding the ZIF-8 self-assembly mechanism via direct experimental techniques.\cite{Cravillon2011,Cravillon2012,Venna2010,Jin2023,Dok2025, Balog2022,Talosig2024,Moh2013} Early on, Venna and coworkers hypothesized that an intermediate amorphous phase was formed during the ZIF-8 synthesis process based on \textit{in situ} XRD and TEM experiments.\cite{Venna2010} There is now wide consensus in the existence of this intermediate species, as it was detected by several other research groups and for other ZIFs as well.\cite{Jin2023,Dok2025, Balog2022,Talosig2024} The nucleation part of the ZIF-8 synthesis process was recently studied by some of us from the computational standpoint\cite{Balestra2022,Balestra2023,Gargari2025} via a model in which the Zn--N bond can be reversibly formed by means of a partially reactive force field we developed called nb-ZIF-FF.\cite{Balestra2022} This methodology allowed us to follow the formation of the first building blocks, their polymerization to form oligomers that then merge and give rise to cyclic structures up to the formation of the amorphous intermediate species. A recent multidisciplinary work has postulated that after the formation of the amorphous intermediate, a reorganization process would lead to the formation of proto-crystals that would then undergo Ostwald ripening to yield the resulting MOF nanoparticles.\cite{Dok2025} Little is known about this last phase of the synthesis process. Even though other computational works have focused on studying the self-assembly of other MOFs,\cite{Yoneya2015,Biswal2016,Colon2019,Kollias2019,Filez2021,Wells2019} none of them has ever focused on studying the growth part of their self-assembly process. The main hurdles that prevent us to model MOF growth are associated to computational cost. On the one hand, the initial system would need to include a pre-formed nuclei, which implies a number of particles too large for relying on \textit{ab initio} approaches, the natural method of choice to treat reactive systems. On the other hand, we would still need to have free reactants available in the solution that is in contact with the nuclei, and we would need their concentration to be steady throughout time to properly take into account the supersaturation condition that direct experiments provide.  

In this work, we overcome these issues by modelling the growth of ZIF-8 at the atomistic level, starting from a pre-formed model proto-crystal in contact with an ionic solution. The concentration of the MOF constituents is kept steady near the proto-crystal surface. Three concentrations of reactants and two temperature regimes are explored. We find that the nature of the species that are adsorbed at the proto-crystal external surface and their preferred adsorption sites strongly depend on the concentration. Increasing either temperature or concentration leads to larger deviations from the crystalline structure as well as to higher growth rates. Compared to nucleation,\cite{Gargari2025}
larger rings are more likely to be formed during the early growth under the conditions of study. The growth process appears to be dominated by adsorption instead of diffusion of the ionic species.

This article is organized as follows. The methodological main details are presented in section II. Our results, subdivided into structure and kinetics, are included in section III. Finally, the conclusions are summarized in section IV.

\section{Computational details} 

\begin{figure}
\begin{center}
\includegraphics[width=0.8\textwidth]{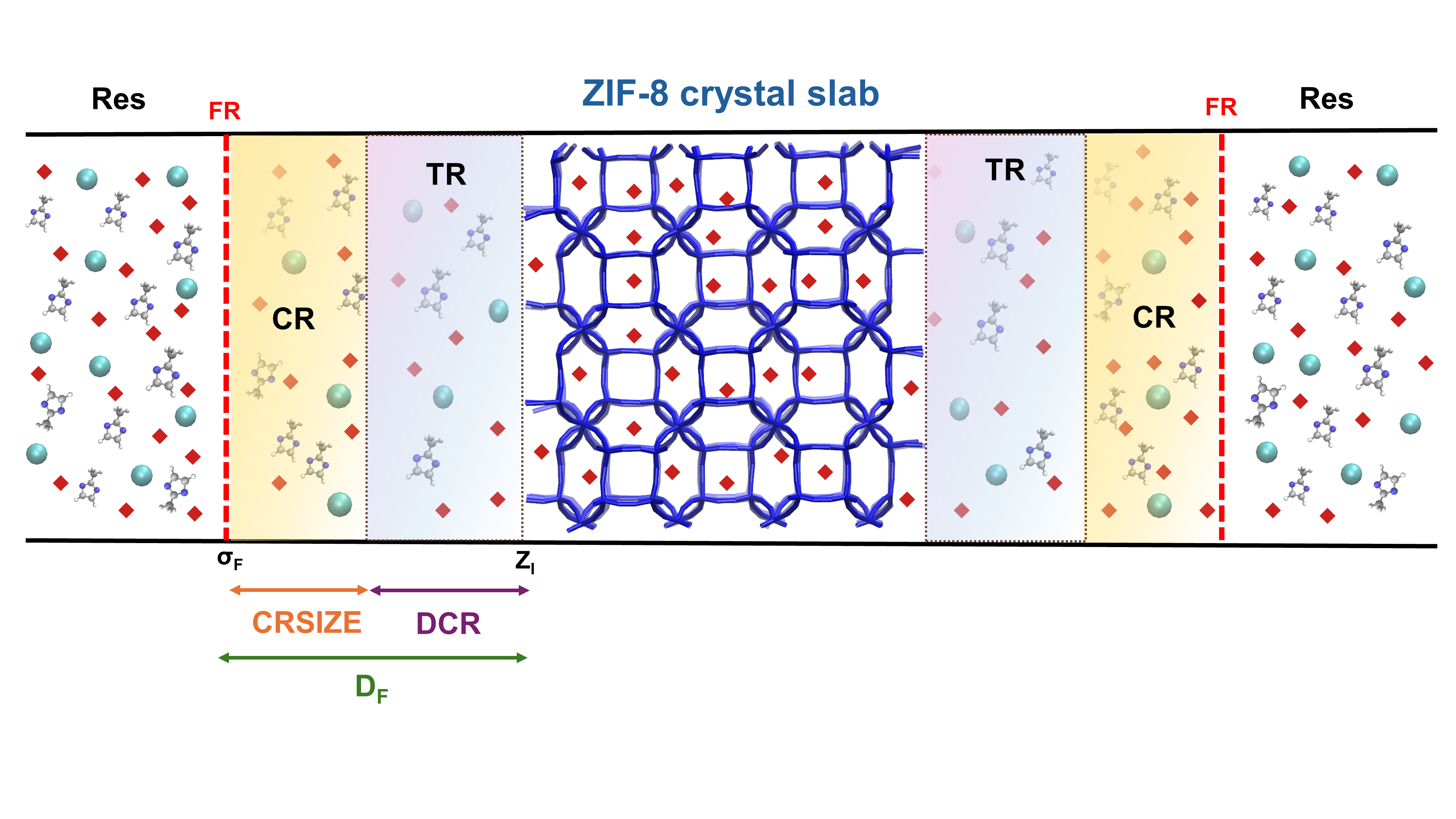}
\caption{Schematic representation of the simulation system showing a symmetric box containing a ZIF-8 proto-crystal slab at its center. Red diamonds represent solvent molecules, cyan spheres represent Zn$^{2+}$ ions, and the ball-and-stick model corresponds to the MIm$^{-}$, with carbon atoms shown in gray, nitrogen in blue, and hydrogen in white. The transition region (TR), control region (CR), and reservoir (Res) are highlighted in transparent purple, orange, and white, respectively. The red dashed line indicates the position where the external force is applied. DCR= size of TRs, CRSIZE = size of the CRs, DF = DCR + CRSIZE.} 
\label{fig:cmumd}
\end{center}
\end{figure}

A mentioned above, the concentration of the reactants is stable within the microscopic timescales in which the growth process takes place at the laboratory, in direct synthesis experiments. However, standard molecular dynamics (MD) simulations are usually performed in ensembles that involve a fixed number of molecules. This limitation leads to a decrease in solute concentration as the crystal grows, which can significantly affect the crystal growth behavior.\cite{salvalaglio2016overcoming,reguera2003phase,grossier2009reaching}
To overcome challenges associated with the depletion of reactants during crystal growth simulations, we combined the constant chemical potential molecular dynamics (C\textmu MD)\cite{perego2015molecular} method with a solute particle insertion technique. In the C\textmu MD method, the simulation box is divided into smaller volume regions, as illustrated in Fig.~\ref{fig:cmumd}. Periodic boundary conditions are imposed along all directions. The ZIF-8 proto-crystal slab is placed at the center of the simulation box and it is exposed to the solution along the \textit{z}-axis. Control regions (CR, shown in orange in Fig.~\ref{fig:cmumd}) are defined on both sides of the proto-crystal slab, where the Zn$^{2+}$ and MIm$^-$ ions number densities are kept constant by applying a force at its farthermost end (force region, FR). The volume separating the crystal slab interface from the CR is referred to as the transition region (TR). This is the region in which the growth will take place, so the physico-chemical behaviour of the molecules and ions that are within it must be correctly described and unaffected by the force that is applied at the FR. The molecular reservoir (Res) is located beyond the FR and ensures the stability of the CR density by supplying Zn$^{2+}$ and MIm$^-$ ions when needed. As an initial step of the C\textmu MD procedure, the Zn$^{2+}$ and MIm$^-$ number densities $(c_t)$ inside the CR are determined as
\begin{equation}\label{eq:1}
c_t = \frac{1}{V^{\rm CR}} \sum_{j=1}^N f(z_j)
\end{equation}
where $V^{\rm CR}$ denotes the volume of the control region, and $z_j$ is the $z$-coordinate of the $j$-th particle. $f(z_j)$ is a smoothed version of a step function that gives a value close to one if $z_j$ is inside the CR and a value close to zero otherwise.
To maintain $c_t$ close to a defined target concentration $c_0$, an external force $F_c$ is applied to the atoms that are close to the boundary between the CR and the Res. The force is expressed as:

\begin{equation}\label{eq:2}
F_c(z) = \kappa (c_t - c_0) G(z)
\end{equation}
where $\kappa$ is the force constant and $G(z)$ is a localized bell-shaped function centered at the CR-Res interface with a width $\sigma_F$ chosen so that the force does not act over the TR. A detailed explanation of Eqs. (1) and (2), along with the definition of $G(z)$, can be found in the work of Perego et al,\cite{perego2015molecular} in which this methodology was first developed.
The C\textmu MD parameters used for all simulations were $\sigma_F$= 0.145 nm, CR= 1.0 nm, TR= 1.8 nm and $\kappa$= 10000 $\frac{nm^3 kJ}{mol}$. The size of the TR was chosen to be larger than the cutoff for dispersion interactions. We selected the values of force width and constant that most efficiently maintained constant concentrations in the CR region. The CR size was selected to be large enough to avoid too large concentration fluctuations and small enough to make the simulations more efficient. Further details on the selection of the C\textmu MD parameters are provided in the Supporting Information (SI). 

Even though the C\textmu MD method helps keep the concentration of reactants constant in the CR for a while, the concentration of reactants in the reservoir naturally decreases over time as species are consumed by crystal growth, which can eventually lead to the depletion of the reservoir and, in turn, of the CR. To avoid this phenomenon, we integrated particle insertion moves to the simulation setup in order to add Zn$^{2+}$ and MIm$^-$ ions into the reservoir when their concentration becomes lower than a threshold value. Specifically, when the number of Zn$^{2+}$ ions in the reservoir region (n$_{Zn^{2+}}$) reaches a value lower than 10, the  C\textmu MD simulation is stopped and Zn$^{2+}$ ions are added in this region until n$_{Zn^{2+}}$ becomes 15. Since the system is dense due to the presence of solvent, these additions cannot be done at random positions, so a metropolis criterion was employed to accept/reject creation attempts.\cite{Metropolis1949} In this way, nonphysical insertions are avoided.
This scheme is implemented via the Grand Canonical Monte Carlo functionality of LAMMPS.\cite{lammps} 
During the insertion attempts, Zn$^{2+}$ and MIm$^-$ moieties remain frozen while solvent molecules are allowed to relax. MIm$^-$ ions were added following the same procedure, but in stoichiometric ratio to ensure that the system remains charge neutral and the metal:ligand ratio is kept fixed. These modifications do not interfere with the dynamics of the central region, since the reservoir is separated from it by the control region. This kind of hybrid MD/particle insertion scheme was already implemented in previous works to reach constant concentration.\cite{kim2023,Goodfellow1996,PrezRamrez2020} More details of this procedure are described in the SI.

The initial configurations were built using PACKMOL\cite{Martnez2009} and they contain a ZIF-8 proto-crystal slab positioned at the center of the simulation box. The solvent consists of 1826 dimethyl sulfoxide (DMSO) molecules and 36 Zn$^{2+}$ and 72 MIm$^-$ ions are initially added in the Res. All systems are then equilibrated under NVT and NPT conditions at T = 298 K and P = 1 bar. To improve statistical reliability, three independent simulations were generated for each system. The systems were modelled at 298 K and differed in their ion concentrations in the CR ($\left[\text{Zn}^{2+}\right]$ = 0.1 M, 0.2 M and 0.4 M). Additionally, simulations at 320 K were performed for the system with a $\left[\text{Zn}^{2+}\right]$ = 0.2 M. In experiments, the concentration used typically spans a broad range from approximately 10$^{-4}$ M to 0.1 M.\cite{manning2023unveiling} For our simulations, we selected the higher end of this spectrum (0.1 M), as well as systems 2 and 4 times more concentrated. Simulating the lower concentrations would have been too computationally demanding because of the prohibitively large system sizes required. ZIF-8 has been synthesized experimentally over a wide solvent-dependent temperature window from ambient conditions up to solvothermal regimes of about 420 K.\cite{Feng2016,Cravillon2012} 

The nb-ZIF-FF force field \cite{Balestra2022} was used for modeling $\text{Zn}^{2+}$ and $\text{MIm}^{-}$ ligands, in which Zn--N interactions are described by a Morse potential allowing reversible coordination bond formation. Simulations were performed in the NPT ensemble, with temperature and pressure regulated using Nosé–Hoover thermostats and barostats,\cite{shuichi1991constant,hoover1985canonical} employing damping constants of 100 and 1000 times the integration timestep, respectively. A timestep of 0.5~fs was used, and long-range electrostatic interactions were evaluated through the Ewald summation method\cite{ewald1921berechnung}  with a relative error tolerance in the forces of 10$^{-8}$. A cutoff of 1.3 nm \ was employed to compute short-range interactions.
The C\textmu MD method is implemented within the PLUMED\cite{Tribello2014} software package, interfaced with LAMMPS\cite{Thompson2022} for all simulations (more information about the implementation is available at the associated \url{https://github.com/mme-ucl/CmuMD}, while the code we used can be found at \url{https://github.com/mme-ucl/PLUMED_Mclass_22-08/blob/main/src/Cmumd.cpp}). Production runs were stopped when the steady state was reached for the growth process. We checked that all simulations reached steady state growth by monitoring the number of ions in the growing phase over time. Steady state is described by a linear growth in time, as will be discussed further below.

Most of the analyses that are performed over the obtained trajectories rely on the identification of the ionic species that are connected directly or indirectly to the central proto-crystal at each simulation step. To perform such assignment, we employed a deep-first search algorithm that identifies clusters of connected components in a graph.\cite{Kozen1992} Ions for which the distance between the carbon atom between the N atoms in the imidazole and the Zn$^{2+}$ is shorter than 0.4 nm are considered to be connected. This cutoff distance is the minimum after the first peak in the corresponding pair distribution function computed for the ZIF-8 crystal. Once the clusters are detected, all the ions that belong to the largest cluster are considered to be part of the growing solid.

\section{Results}

\subsection{Structure of the grown layers}

\begin{figure}[H]
   \centering
    \includegraphics[width=0.9\textwidth]{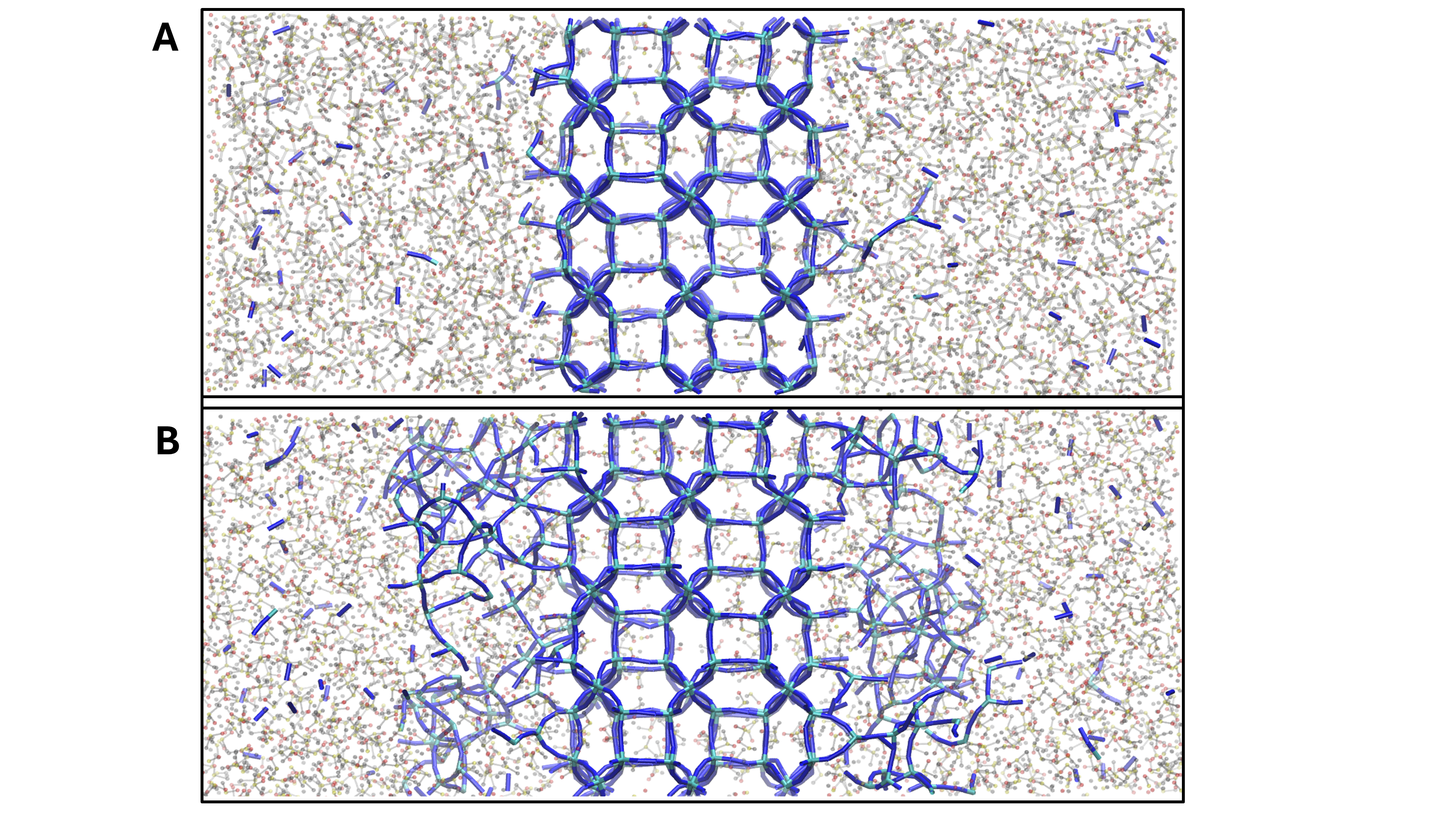}
    \vspace{0.1cm}
    \caption{Snapshots illustrating (A) the initial and (B) the final configurations of one of the independent simulations for the  [Zn$^{2+}$]= 0.4 M and T=298 K system. Zn$^{2+}$ ions are shown as cyan spheres, ligands are represented as blue lines connecting their two nitrogen atoms, DMSO molecules are displayed with carbon in grey, hydrogen in white, oxygen in red and sulfur in yellow.}
    \label{fig:growth_mechansim}
\end{figure}

In the simulated systems, the layers that correspond to the growth of the proto-crystal exhibit different degrees of disorder as the thermodynamic conditions change. This agrees with the known fact that both temperature and concentration of reactants can tune the degree of crystallinity or amorphousness in MOF synthesis.\cite{Shaw2024,Han2022,Allegretto2024}
Fig.~\ref{fig:growth_mechansim} illustrates a representative example of disordered growth, showing the initial and the final configurations of the production run for one of the independent simulations of the [Zn$^{2+}$]=0.4 M at T=298 K system, which corresponds to the highest concentration studied. The amorphous nature of the region that results from the growth can be clearly observed in panel B of the figure.

The visual inspection of the trajectories shows that ions entering the TR can associate and form short, mostly linear oligomeric units that subsequently migrate toward the surface and are adsorbed. These kind of structures were observed more frequently in the high concentration systems and are known to be stable.\cite{Mendez2025,Gargari2025}
We support this statement by computing the fraction of oligomers of size N formed in the TR at each of the conditions studied. The results are plotted in Fig. S2 of the SI, and provide quantitative evidence of the higher proportion of oligomers in the high concentration setups as observed by visual inspection.
In all cases, once the ion or oligomer is adsorbed at the surface, local rearrangements occur that lead to the formation of ramified motifs or rings. 

\begin{figure}[H]
   \centering
    \includegraphics[width=0.5\textwidth]{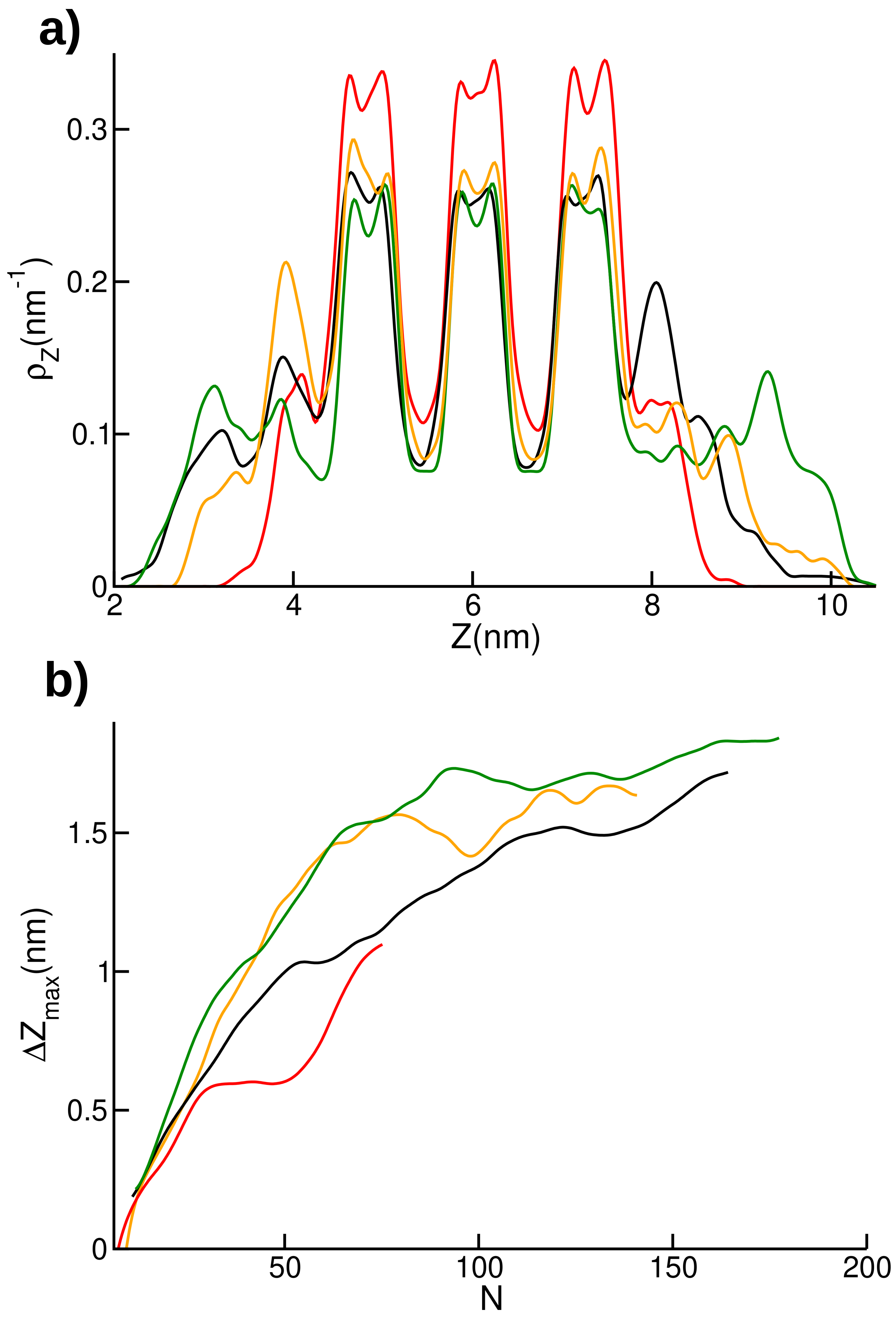}
    \vspace{0.1cm}
    \caption{(a) Ionic density as a function of the $z$ coordinate for each of the systems studied at the end of the simulation. Only ions connected to the central cluster are considered. One out of the three independent simulations is shown in each case. The densities are normalized to integrate to one. (b) Maximum $z$-distance of the grown layer with respect to the original proto-crystal slab as a function of the total number of ions that were added ($N$). Each plot corresponds to an average over six surfaces (three independent simulations that contain two external surfaces each). Color code: [Zn$^{2+}$] = 0.1 M, T = 298 K (red), [Zn$^{2+}$] = 0.2 M, T = 298 K (black), [Zn$^{2+}$] = 0.4 M, T = 298K (green) [Zn$^{2+}$] = 0.2 M, T = 320 K (orange). }
    \label{fig:density_z}
\end{figure}

In order to gain further insights into the nature of the growth process, we studied the effect of concentration and temperature in the growth along the $z$ direction (perpendicular to the proto-crystal plane). Panel (a) of Fig.~\ref{fig:density_z} shows the linear density profile of the ions attached to the surface at the end of a simulation for each system. All four plots contain three central peaks that correspond to the ions that conform the proto-crystal initially placed in the simulation box as well as additional peaks outside the central region that correspond to the ions attached during the growth simulation. 
It is worth to mention that since each simulation reaches a different number of ions attached, the intensity of the central peaks differs between plots as a consequence of normalization. The results from the other independent simulations present qualitatively similar behaviours.
Even thought the growing phases are not crystalline as previously discussed, the fact that the density profiles present quasi-equidistant peaks indicates that some structural features of the proto-crystal seed are inherited by the newly grown phase. This behaviour is more marked in the plot that corresponds to ([Zn$^{2+}$]= 0.2 M, T = 298 K)  conditions (black curve), in which we can observe two peaks at the left- side and one at the right-hand side of the proto-crystal peaks that can be associated with new layers of the material. The curve that corresponds to the [Zn$^{2+}$]= 0.4 M condition (depicted in green) presents a much more flat profile, which may be a consequence of the increase in supersaturation, which could favor disordered growth.
A similar, albeit less marked, tendency is found for the simulations at 320 K (orange curve). It is known that an increase in the diffusion coefficient of the solute (in this case because of the temperature rise) leads to a more disordered growth.\cite{Liu2023} 
The results from the most diluted system (red curve) present less amount of peaks due to the fact that in these conditions the growth is slower, so less layers were formed within the simulation time. Still, two emerging peaks can be identified. This system shows more symmetry in terms of left- and right-hand external surface additions and, following the previous arguments, it is expected to present a more ordered growth. We will confirm this hypothesis in our subsequent analyses.  

The maximum distance between the original proto-crystal slab and the grown layer as a function of the amount of added ions is plotted in the panel (b) of Fig.~\ref{fig:density_z}, with the aim of obtaining a more quantitative picture of the degree of disorder in the grown layers for all systems. These results correspond to an average over six surfaces (three independent simulations, each one containing two external surfaces of the proto-crystal). Both Zn$^{2+}$ and MIm$^-$ ions are considered in this analysis in an equal foot, as the ratio of ions added to the surface follows almost perfectly the 1:2 stoichiometry of the crystal (we will discuss this further later on), and no distinct features arise from the plots that correspond to individual ions. The advantage of plotting the results as a function of the amount of added ions instead of time is that we can compare results from simulations that grew at different rates. We can observe that for the same number of added ions N, the maximum distance from the original proto-crystal slab is shorter in the diluted case (red curve) than in the others, followed by the intermediate concentration, ambient temperature case (black curve). Results that correspond to the higher temperature or higher concentration systems present the largest values (green and orange curves). This is an indicator of how uniform the new built layers are, since a higher value of distance means that the new ions were added in a more dendrite-like fashion. These results support the previous argument that states that an increase in concentration or in temperature produces a more disordered growth. It may be the case that the disorder we see in our simulations is correlated to the fact that the concentrations that we are probing are on the upper limit of those employed at the laboratory, as mentioned above. This is in line with our observation that the lower concentration investigated leads to a more ordered growth. 

\begin{figure}[H]
   \centering
    \includegraphics[width=0.8\textwidth]{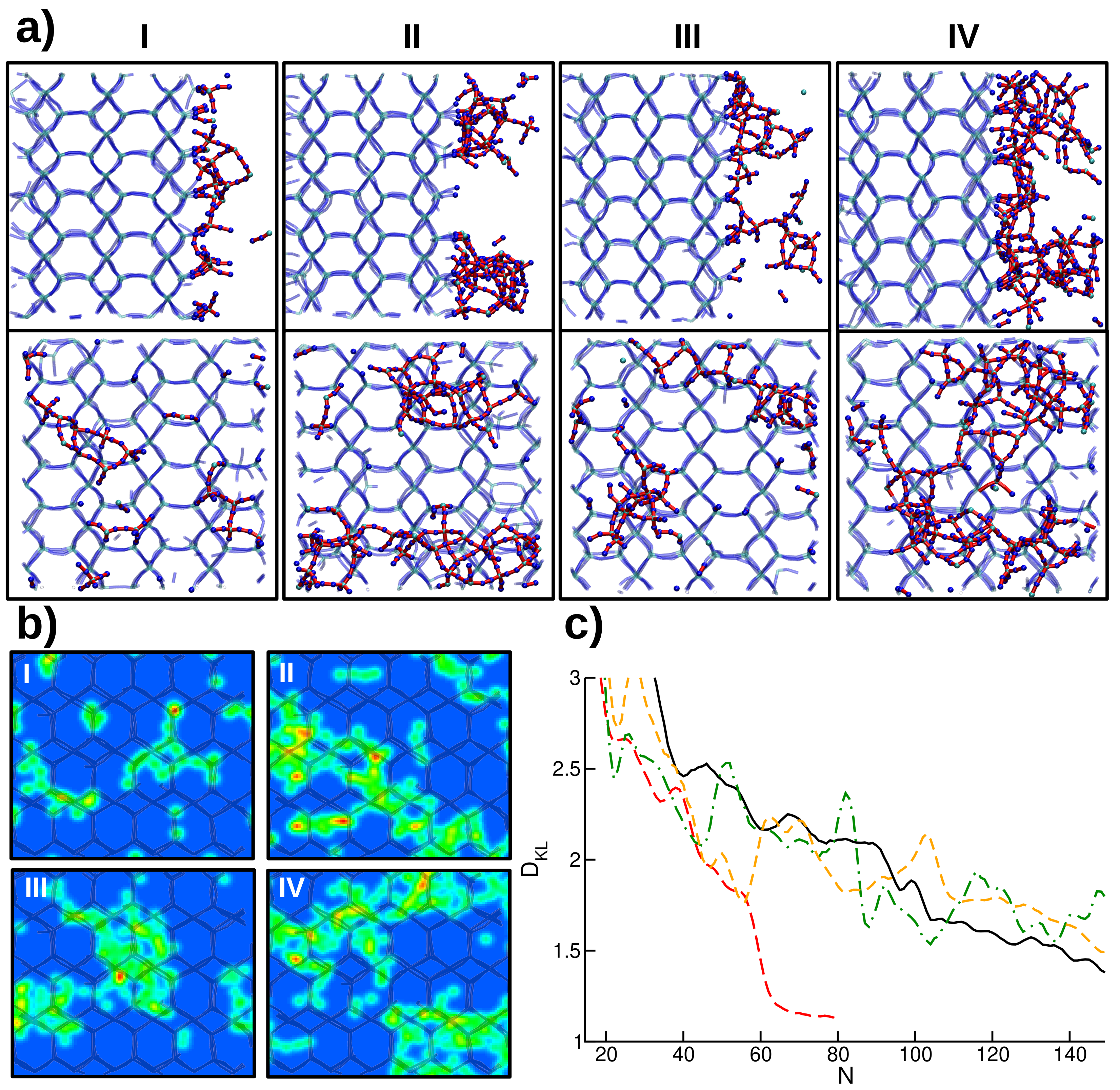}
    \vspace{0.1cm}
    \caption{a) Snapshots of the growth on the ZIF-8 proto-crystal surface at several ion concentrations and temperatures. Red sticks indicate newly formed bonds, cyan spheres show newly coordinated Zn$^{2+}$ ions, and blue spheres represent the N atoms of newly attached ligands. The underlying crystal surface is shown in the background as blue sticks and cyan balls. The uppermost plots depict a side view of the surface while the rest show the top view for each system. 
    b) Snapshots of the surface density of the added ions in the $xy$ plane at the end of one of the independent simulations for each system. The density color scale goes from blue to red as the density increases. Densities are normalized to integrate to one. Throughout  the figure, each system is noted as (I) [Zn$^{2+}$] = 0.1 M, T = 298 K, (II) [Zn$^{2+}$] = 0.2 M, T = 298 K, (III) [Zn$^{2+}$] = 0.2 M, T = 320 K and (IV) [Zn$^{2+}$] = 0.4 M, T = 298 K.
    c) Kullback–Leibler divergence between the ion surface density distribution and the reference distribution as a function of the amount of ions (N) added to the surface. Each curve corresponds to an average of the six simulated surfaces. Systems (I), (II), (III) and (IV) correspond to red, black, orange and green curves respectively.}
    \label{fig:2D}
\end{figure}

We continue our description of the growth process with the analysis of the in-plane growth of the solid grown on the proto-crystal surface. Example snapshots of the growth on the ZIF-8 proto-crystal surface are presented in panel (a) of Fig.~\ref{fig:2D} for all systems studied. The uppermost snapshots depict a side view (perpendicular to the surface), highlighting the extension of grown layers in the $z$ direction, while the other ones show a top view, in the $xy$ plane (parallel to the surface), revealing the lateral arrangement of rings on the surface. At low concentration the growth remains very limited: only a few ions attach to the exposed surface sites, forming a single short linear chain and a few rings. At intermediate concentration, more rings appear in different parts of the surface, producing a disordered and defective, amorphous-like layer. At higher concentration and temperature, the surface becomes highly irregular and branched, with dendritic features. This behavior indicates a transition to rough growth, where the attachment of ions or oligomers is fast and no longer follows the original proto-crystal shape.

To obtain a more quantitative description of the distribution of the added atoms in the $xy$ plane, we found it useful to transform the discrete set of ionic coordinates into a continuous density distribution function. To do so, we defined the distribution function as a sum of Gaussian functions centred in the ion positions:

\begin{equation}
\rho(x,y) =\frac{1}{Q} \sum_i e^{-\frac{(x-x_i)^2+(y-y_i)^2}{2\sigma^2}}
\end{equation}
where $\sigma$ represents the Gaussian width (0.1 nm) and $Q$ is a normalization constant that makes the distribution integrate to one over the full plane. The sum includes the ions $i$ that belong to the central cluster but lie outside the initial proto-crystal region, so that only the added species are considered. This is done by filtering the ions by their position in the direction perpendicular to the proto-crystal external surface ($z_i$ coordinate). For each simulation two density plots are obtained: one for the left-hand external surface ($z_i$ $<$ 4 nm) and one for the right-hand one ($z_i$ $>$ 8 nm).
In Fig.~\ref{fig:2D} (b) we show typical snapshots of the surface distribution functions for each of the studied systems at the end of the simulation. To show the diversity of the formed patterns, the simulations depicted in this plot are not the same ones shown in the (a) panel, but independent analogues.
In each figure we superimposed a picture of the underlying proto-crystal slab, to identify if the added ions follow the shape of the crystal or not. It is clear that in the diluted system (marked as I in the figure), the added atoms are located in positions that resemble the ones of the crystal structure. In addition, it is more likely for a new ion to be inserted on top of the initial surface than to attach over the newly grown layer. 
On the other hand, the rest of the systems show a lower correlation with respect to the template positions, although they are not completely random either. The probability of the ions being added on top of the newly grown layer instead of filling the holes on top of the initial surface increases.
To carry out a more statistically grounded analysis, we quantified the divergence between the obtained distributions and the distribution of the perfect crystal $\mathcal{P}_0(x,y)$, through the calculation of the Kullback–Leibler divergence $D_{KL}$\cite{Gianfelici2009}:

\begin{equation}
    D_{KL} = \int \rho(x,y) ln\frac{\rho(x,y)}{\mathcal{P}_0(x,y)} \mathrm{d}x \mathrm{d}y
\end{equation}
$D_{KL}$ is always a positive number that will decrease if $\rho(x,y)$ gets closer to the reference distribution, and will be exactly zero if they are equal. In panel (c) of Fig.~\ref{fig:2D} we plotted the value of $D_{KL}$ as a function of the number of added ions N averaged over the six surfaces, in the same spirit as in panel (a) of Fig.~\ref{fig:density_z}. 
From the plot we can observe that the diluted system presents the smallest $D_{KL}$ values, meaning that the added ions are distributed following the prot-crystal pattern. This tendency was also identified in the analysis of the growth in the $z$ direction shown in Fig.~\ref{fig:density_z}, in which we identified a more ordered growth in this system with respect to the others. The other three curves present comparable values, although the [Zn$^{2+}$] = 0.2 M, T = 298 K system yields lower $D_{KL}$ values for high N, in agreement with our previous analyses.

\begin{figure}[H]
    \centering
    \includegraphics[width=0.95\textwidth]{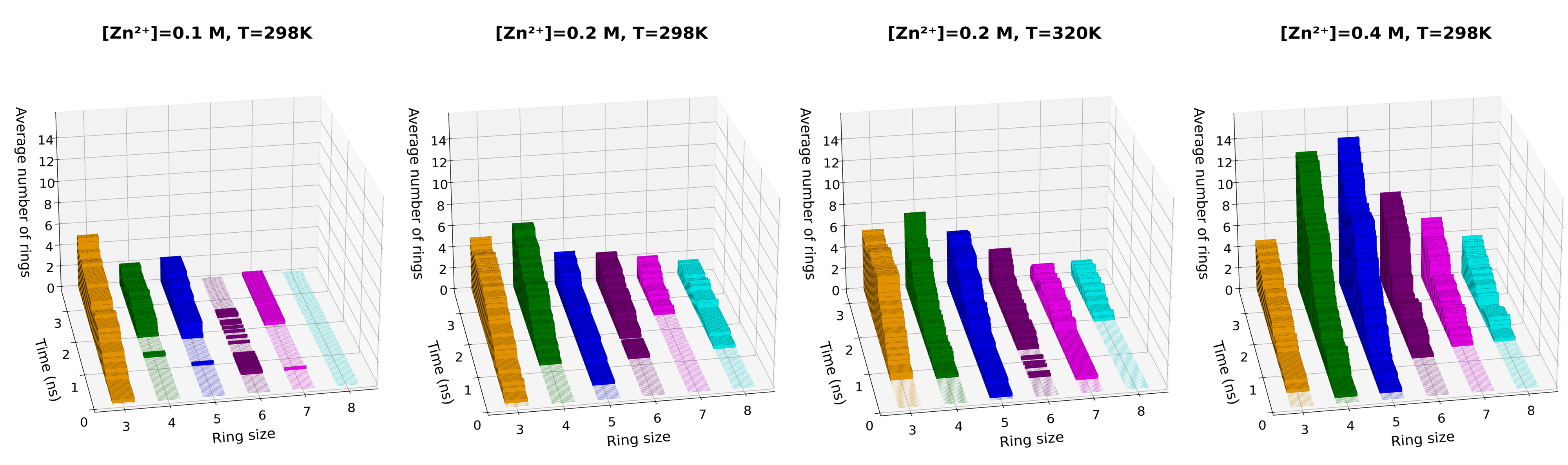}
    \vspace{0.1cm}
    \caption{Time evolution of the average number of rings of different sizes ranging from 3 to 8 Zn$^{2+}$ ions for systems with concentrations of ([Zn$^{2+}$]= 0.1, 0.2 and 0.4 M at T=298 K as well as for the system with [Zn$^{2+}$]= 0.2 at T=320 K. Each data point represents the average from three independent simulations. }
    \label{fig:ring_synthesis}
\end{figure}

Rings are structural motifs usually employed to describe MOFs and other porous materials.\cite{castel2022challenges, franzblau1991computation,o2012deconstructing,hobday2018understanding} These are cyclic structures formed by alternating metal--ligand bonds, typically defined according to the number of metal ions that compose them. To complete our structural analysis, we studied the ring distribution in the grown layer. Bulk ZIF-8 contains 4 and 6-membered rings (MR).
Fig.~\ref{fig:ring_synthesis} shows the evolution of the number of rings of various sizes as a function of simulation time for all systems ([Zn$^{2+}$]= 0.1, 0.2, 0.4 M at 298 K
and [Zn$^{2+}$]= 0.2 M at 320 K), averaged over all three independent simulations in each case. We analyzed the ring evolution over the first 4~ns. It can be seen that increasing ion concentration and temperature strongly affects the evolution of ring structures.
 
At the lowest concentration ([Zn$^{2+}$]= 0.1 M, T= 298 K) ring formation proceeds slowly, resulting in only a small average number of 4- and 5-MR rings. 3-MR rings are most frequently observed under these conditions, while 6-MR rings are not stabilized and 8-MR rings are not formed during the first 4 ns of simulation. However, when the concentration is doubled ([Zn$^{2+}$]= 0.2 M, T= 298 K), a noticeable increase in the formation of 4-MR rings is observed, making them the dominant species, while 6-, 7- and 8-MR rings also form albeit in low numbers. Moreover, we examined the effect of temperature at this concentration to determine how increasing the temperature influences ring formation. By increasing the temperature at the same concentration ([Zn$^{2+}$]= 0.2 M, T= 320~K), we found that higher temperature promotes the formation of 3-, 4-, 5-, and 6-MR rings, while the 4-MR rings remain the predominant species, as observed in ambient temperature conditions. At higher temperature, faster ion motion increases the chances of forming 5-MR rings, allowing them to form faster than the other rings. At the highest concentration studied ([Zn$^{2+}$]= 0.4 M, T= 298 K), all rings become more abundant. The increase in 4-, 5- and 6-MR ones is the most significant. 5-, 7- and 8-MR rings likely represent temporary or defect-related structures that will later rearrange into 4- and 6-MR rings yielding the ordered ZIF-8 framework. 

\begin{figure}[H]
    \centering
    \includegraphics[width=0.8\textwidth]{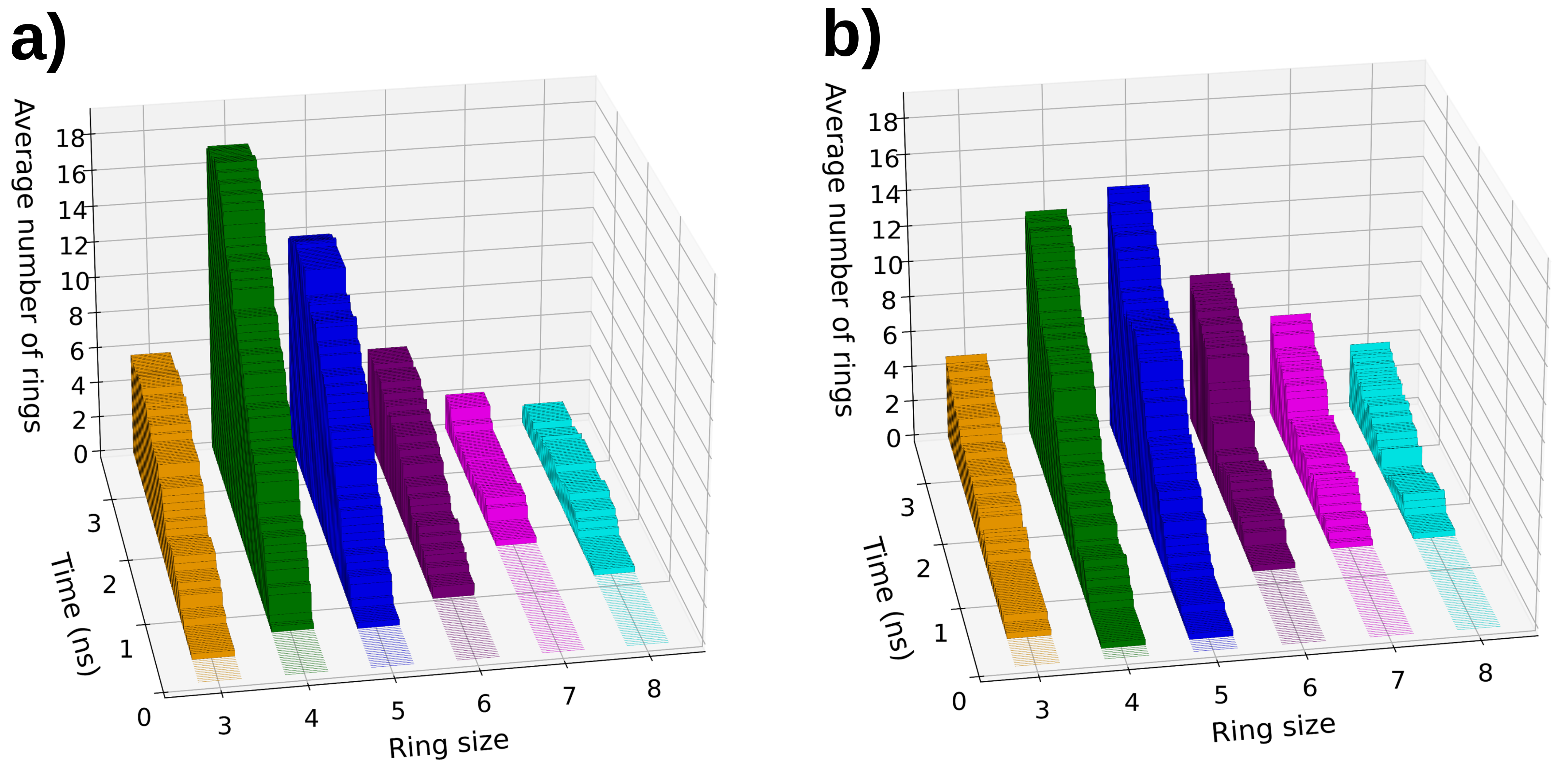}
    \vspace{0.1cm}
    \caption{Comparison of the ring structural evolution during ZIF-8 (a) nucleation and (b) growth. T = 298 K and [Zn$^{2+}$]= 0.4 M or 0.53 M for growth (this work) and nucleation (Ref.\cite{Gargari2025}), respectively. Each data point represents an average over three independent simulations.}
    \label{fig:ring_nuc_growth}
\end{figure}

To compare the structural evolution during nucleation and growth, we analyze systems with similar temperature and comparable concentrations. The nucleation of ZIF-8 was investigated in a previous work carried out in our group at [Zn$^{2+}$]=0.53 M and T=298 K.\cite{Gargari2025} In that work, simulations started from the solvated Zn$^{2+}$ ions and MIm$^-$ ligands, which formed chemical bonds over time, finally leading to an amorphous phase, which also contained a variety of rings sizes. Fig.~\ref{fig:ring_nuc_growth} presents a comparison between the ring distribution (a) in the amorphous phase formed during nucleation and (b) that formed during growth at the same temperature and comparable ion concentration [Zn$^{2+}$] = 0.4 M. In the growth case, 4- and 5-MR rings start forming faster. The early formation of 3-MR rings is common both to nucleation and growth. The higher average number of 6-MR rings in the growth regime indicates that the system is progressing toward the ZIF-8 crystalline framework. Moreover, the higher average number of 5-, 7-, and 8-MR rings observed in the growth compared to the nucleation suggests the presence of transient structural motifs, which may reorganize into the more stable 4- and 6-MR rings forming the sodalite topology as the structure continues to evolve. Overall, it is observed that nucleation is crucial in the formation of 4-MR rings, while the formation of 6-MR rings seems to be more favored during growth. 

Our results suggest that in this range of concentrations, the growth mechanism involves the formation of an amorphous interface, which could undergo a subsequent rearrangement to give rise to a completely crystalline phase at timescales longer than typical simulations times. 
Nevertheless, the crystalline features of the formed layers augment significantly for the lowest concentration studied, suggesting that a further decrease in concentration would create the conditions for ordered growth. It would be desirable to run simulations at lower concentrations. We plan to integrate enhanced sampling techniques within these schemes in future work to make this possible, as previously done to model the growth of organic solids.\cite{Karmakar2018}

\subsection{Growth rates}

\begin{figure}[H]
   \centering
    \includegraphics[width=0.5\textwidth]{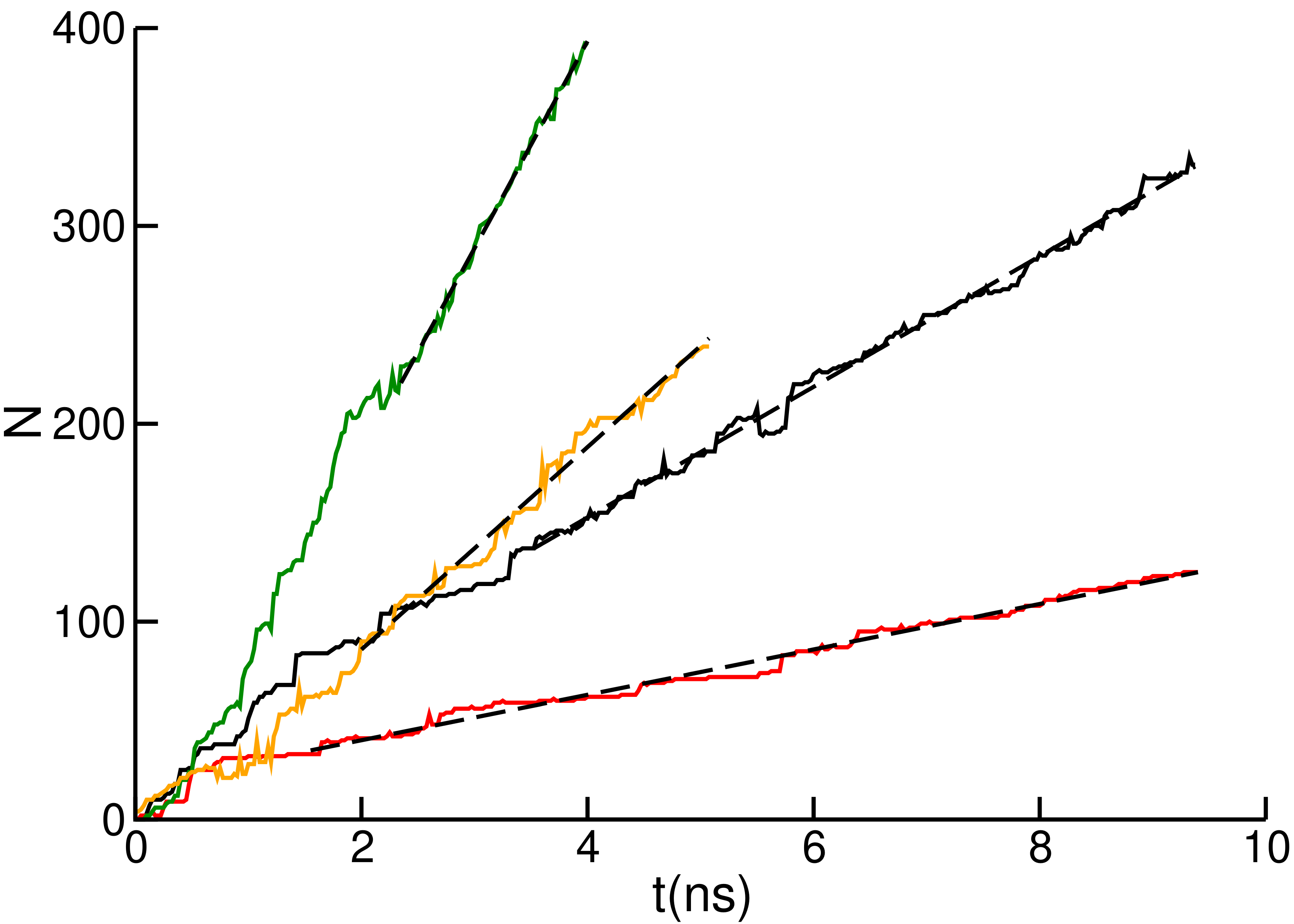}
    \vspace{0.1cm}
    \caption{Number of ions added to the central cluster ($N$) as a function of time for simulations with [Zn$^{2+}$] = 0.1 M, T = 298 K (red), [Zn$^{2+}$] = 0.2 M, T = 298 K (black), [Zn$^{2+}$] = 0.4 M, T = 298 K (green) and [Zn$^{2+}$] = 0.2 M, T = 320 K (orange). Black dashed lines correspond to linear regressions performed over the stationary growth regions of each plot.}
    \label{fig:growth_rates}
\end{figure}

We now turn to the analysis of the kinetics of the growth process. To quantify the growth of the crystal slab, we computed its rate for each of the systems considered. To do so, we monitored the amount of Zn$^{2+}$ and ligands that are part of the largest cluster, which includes the central proto-crystal slab, as a function of time. In Fig.~\ref{fig:growth_rates} we plotted the results for one independent simulation for each system as an example, other results are presented in the SI. In all cases we can identify an initial transient period followed by a stationary growth regime in which the slope of the curve is almost constant. By performing a linear regression (dashed black lines) over the curves in the stationary growth region, we extracted the growth rates.

The same procedure was done for all the independent simulations, this time distinguishing the rates for Zn$^{2+}$ and ligand additions separately. The results, with their respective error values, are summarized in Table \ref{tabla}. By looking at these data, we can first note that the growth rate ratio between Zn$^{2+}$ and ligand is 1:2 in all cases, following the stoichiometric relationship of the crystal. This means that despite being amorphous, the formed surfaces are electrically neutral. Moreover, the error bars of the obtained rates are in the order of 10\%, meaning that the results are consistent within the different independent simulations.

By comparing the rate at different concentrations and equal temperature, it is clear that an increase in concentration leads to an increase in the growth rate as expected, since the supersaturation acts as driving force of the process. 
We also observe that the relationship between the growth rate and the bulk concentration is not linear, which is typically a sign of a growth process controlled by ion adsorption, contrary to the linear relationship expected in pure diffusion controlled processes.\cite{Andreassen2016} This hypothesis agrees with results from previous works that identify the Zn--ligand bond formation as an activated process.\cite{Mendez2025,Gargari2025} The activation energy associated with this reaction corresponds roughly to the Zn--solvent dissociation that precedes the Zn--ligand bond formation and it is in the order of 8 $k_b T$. The non linear behaviour of the growth rate could also be induced by the formation of oligomers in the solution, since classical theories normally assume that growth mechanisms are driven only by monomer attachment.\cite{DeYoreo2003} Nevertheless, several other solids such as calcite and brucite exhibit non classical behaviour, in which oligomer attachments play an important role.\cite{putnis2021} A more systematic study of the concentration dependence of the growth rate would be necessary to fit the data with a model. 
Finally, by comparing the rate at different temperatures and equal concentration, we observe that an increase in T leads to an increase in the rate. Although a higher temperature implies a lower driving force due to the increase in solubility, this is compensated by the simultaneous increase in the rate of the activated processes involved in the growth mechanism, specifically, diffusion and ion attachment. The latter effect seems to overlay the change in solubility under the studied conditions.

\begin{table}[hbtp]
\small
  \caption{Growth rates per unit of surface area for Zn$^{2+}$ and ligands at each of the studied thermodynamic conditions. The error values correspond to the standard deviation between the three independent simulations for each system.}
  \label{tabla}
  \begin{tabular*}{0.7\textwidth}{@{\extracolsep{\fill}}llll}
    \hline
  [Zn$^{2+}$] (M) & T (K) & $v_{\mathrm{Zn^{2+}}}$(ns$^{-1}$nm$^{-2}$)  & $v_{\mathrm{MIm^-}}$(ns$^{-1}$nm$^{-2}$) \\
    \hline
0.1  & 298    & 0.13 $\pm$ 0.01      & 0.25 $\pm$  0.04  \\
0.2  & 298    &   0.43   $\pm$  0.03 &    0.87  $\pm$ 0.05   \\
0.4  & 298    &    1.49  $\pm$  0.1 &       3.03  $\pm$ 0.3     \\
0.2  & 320    &    0.70   $\pm$ 0.05   &       1.33  $\pm$ 0.2  \\
    \hline
  \end{tabular*}
\end{table}

\section{Conclusions} 

In this study we have applied the recently developed  C\textmu MD method in combination with a particle insertion scheme to study the solvothermal growth process of a metal-organic framework from a computational standpoint for the first time. In particular, we studied the growth of the archetypal MOF ZIF-8 spanning a series of concentrations of the reactants and two synthesis temperatures. 

We found that the growth mechanism involves the attachment of both monomers and oligomers. The oligomer formation is more relevant at higher concentrations.
The growth process leads to an interface with amorphous characteristics, rather than forming an ordered sodalite structure. Despite being amorphous, the newly formed MOF layers still exhibit a density profile compatible with the presence of pseudo-pores. At lower concentrations an ordered growth is observed, while at higher concentrations the formed structures present dendritic features.

After the reorganization of Zn--ligand bonds, rings of a variety of sizes are formed within the newly grown layers. In the early evolution of the structure, clear differences in dominant ring sizes appear as concentration and temperature change. At low concentration, 3-membered rings are the most common. When the concentration is doubled, the dominant ring size increases to 4-membered, regardless of the temperature being 298 or 320 K. At higher concentration, 5-membered rings dominate. We can also see that the rings distributions change when increasing concentrations or temperature and the formation of larger rings becomes more likely. When comparing the ring distribution evolution in the growth stage with the nucleation stage that was studied in previous work,\cite{Gargari2025}
we observe distinct trends. Growth promotes the formation of larger rings, which leads to an increase in 6-membered rings, whereas nucleation tends to form smaller rings, with a significant predominance of 4-membered rings. Our results consistently show that increasing concentration or temperature promotes more disordered surface growth and reduces the symmetry of the formed layers on the both sides of the surface. In addition, the ions that are attached to the surface at the lowest concentration investigated better adapt to the surface structure, while higher concentrations and higher temperature lead to larger deviations. It is less likely for an ion to attach over the newly grown layers than over the initial proto-crystal external surface when the concentration is lower.

We have also analysed the growth rate under steady state conditions and found that it increases with concentration and temperature. These results are expected, as the supersaturation drives the growth. The non-linear dependence of the growth rate on concentration suggests that the process is controlled by adsorption rather than diffusion, which is consistent with earlier studies from our group \cite{Gargari2025,Mendez2025} that highlight that the Zn--ligand bond formation is an activated step. Additionally, we have observed that the growing surface remains electrically neutral with a Zn$^{2+}$:ligand ratio of 1:2. 

This work not only delivers new insights on the molecular-level mechanism of the self-assembly of ZIF-8, but also presents a new computational method that can be applied to investigating the growth of any desired MOF under a variety of synthesis conditions. We hope that it will be widely used in the community to further elucidate MOF growth mechanisms.  

\section*{Conflicts of interest}
There are no conflicts of interest to declare.

\section*{Data availability}

The data supporting this article have been included as part of the Supporting Information files or at github \url{https://github.com/rosemino/MAGNIFY/tree/main/ZIF-8_Growth_CmuMD}.

\begin{acknowledgments}
This work was funded by the European Union ERC Starting grant MAGNIFY, grant number 101042514. This work was granted access to the HPC resources of IDRIS under the allocation A0170915688 made by GENCI.
\end{acknowledgments}


\newpage

\MakeTitle{
        Supporting Information for: \\Computer Simulation of the Growth of a Metal-Organic Framework Proto-crystal at Constant Chemical Potential
    }{
        Sahar Andarzi Gargari, Emilio Méndez and Rocio Semino           
    }{
    Sorbonne Université, CNRS, Physico-chimie des Electrolytes et Nanosystèmes Interfaciaux, PHENIX, F-75005 Paris, France
    }

\section*{Optimisation of the C\textmu MD setup}

We tested box lengths ranging from 12 to 20~nm in the $z$ direction (perpendicular to the proto-crystal slab) and examined both three-pore-layer and four-pore-layer non-defective ZIF-8 slab as proto-crystal models. The results were comparable, so we kept the 12~nm model for computational efficiency. The proto-crystals were filled by DMSO molecules in all cases. We explored a wide range of C\textmu MD parameters, varying:
\begin{itemize}
    \item Transition Region (TR) size: 1.0--1.8~nm
    \item Control Region (CR) size: 1.0--2.5~nm
    \item Force width: 0.07--0.7~nm
    \item Force constant: 1000--80000~nm$^3$\,kJ\,mol$^{-1}$
\end{itemize}
We initially tested introducing a concentration gradient inspired by the cannibalistic C\textmu MD method presented by Karmakar \textit{et al.}~\cite{karmakar2018cannibalistic}, with concentrations of 0.33--0.50~M at the CR on the left and of 0--0.083~M at the CR on the right to study growth on the left-hand external surface of the ZIF-8 proto-crystal and dissolution on the right-hand one. The dissolution process was intended to provide the ions required for the growth by means of the reservoir, which connects both regions through the periodic boundaries. However, we found that the dissolution process of the right-hand external surface of the MOF proto-crystal was extremely slow in this setup, and thus the replenishment of the reservoir was too slow to provide ions for the growth within a reasonable timescale. This was also the case even when an additional force was applied to accelerate dissolution. For this reason, we did not continue with this method. Based on all the tests performed, we selected the parameter set that resulted in the most stable concentrations at the CR regions while yielding reasonably fast growth. In our final setup, we thus opted for combining the original version of the C\textmu MD method with a particle insertion algorithm to study growth at both sides of the crystal as in the original work of Perego and collaborators.\cite{perego2015molecular}

\section*{C\textmu MD performance}\label{details}

To evaluate the performance of the C\textmu MD method\cite{perego2015molecular}, we plotted the concentration of metal ions and ligands in each region of the simulation box as a function of time in Fig.~\ref{fig:concentrations_time}. 
It is clear that the concentration is kept constant in the CR (middle panel) apart from thermal fluctuations, so we can conclude that the method is working well. Since the growth of the solid slab occurs in the TR, it is expected that the amount of ions in that region grows as the simulation evolves, which is also the case (upper panel). Finally, to probe the performance of the particle addition scheme, we show that the concentrations in the reservoir fluctuate around fixed values, avoiding depletion phenomena. All independent simulations show similar behaviour.

\setcounter{figure}{0}
\renewcommand{\thefigure}{S\arabic{figure}}

\begin{figure}[H]
   \centering
    \includegraphics[width=0.5\textwidth]{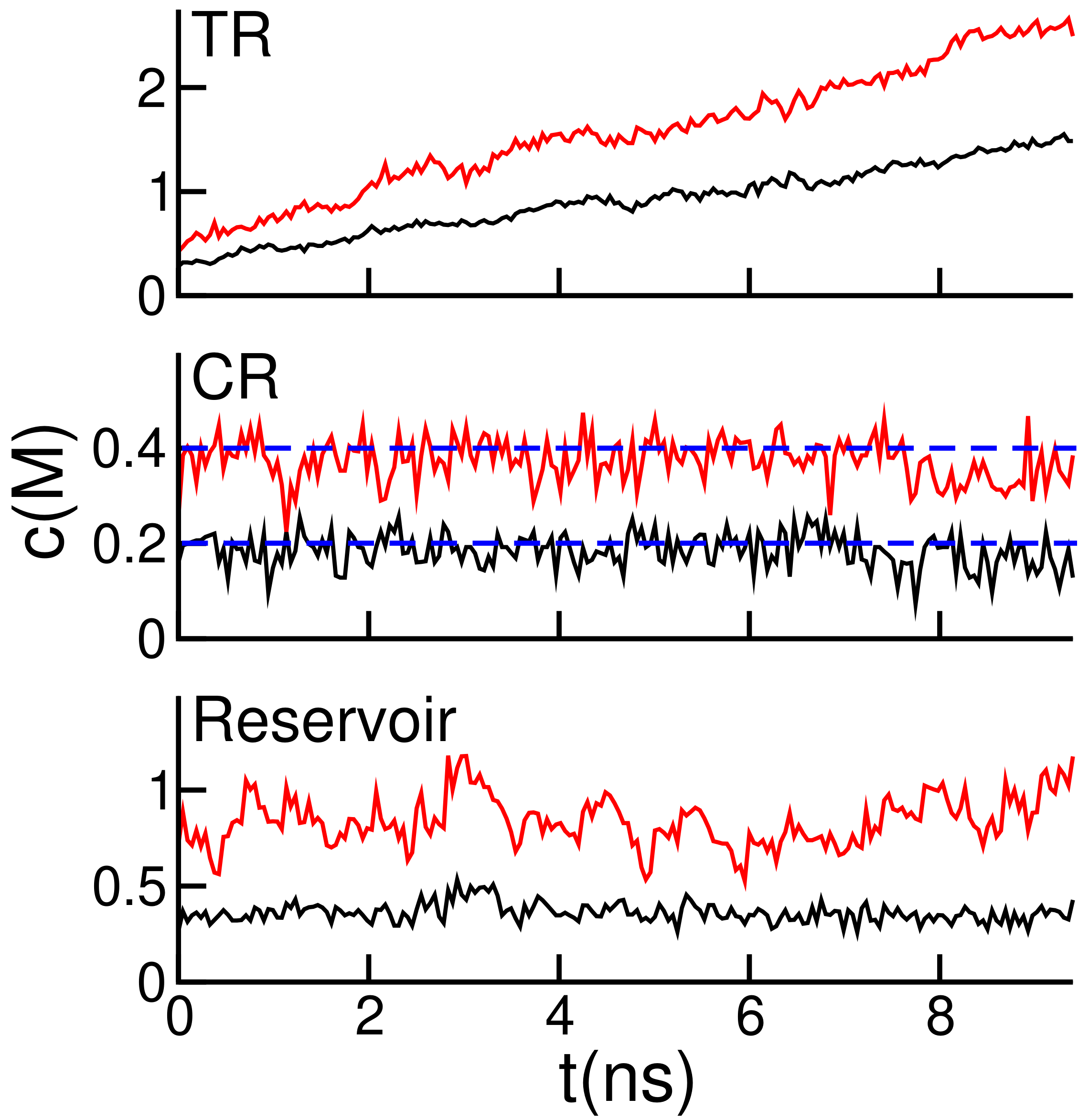}
    \vspace{0.1cm}
    \caption{Concentration vs. time for Zn$^{2+}$ (black) and 2-methylimidazolate (red) in each region of the simulation box. Upper panel: transition region, middle panel: control region, lower panel: reservoir.
    The plots correspond to a simulation with target concentration of Zn$^{2+}$ and ligand of 0.2 M and 0.4 M respectively and T =
    298 K. These values are marked with dashed blue horizontal lines in the control region plot.}
    \label{fig:concentrations_time}
\end{figure}

\section*{Particle insertion method}\label{particle_insertion}

The particle insertion algorithm works by the following set of instructions:
(i) The C\textmu MD simulation runs for 10000 time steps.
(ii) The amount of Zn$^{2+}$ ions in the reservoir is then counted.
(ii) If the number of Zn$^{2+}$ is lower than a threshold value of 10, the system will undergo a set of addition attempts until the Zn$^{2+}$ number reaches an upper threshold value of 15.
(iv) A similar protocol will be applied to add exactly twice the amount of ligands in order to maintain the global stoichiometry.
(v) The previous steps are repeated until the steady state is achieved for the growth (measured by computing the growth rate and verifying it is linear with time).

The addition attempts consist in Monte Carlo moves in which a particle can be added or deleted according to a Metropolis acceptance criterion.\cite{Metropolis1949} This is conceptually equivalent to placing the system in contact with a bigger reservoir of chemical potential $\mu$. The value of $\mu$ is chosen to be large enough to produce a non-negligible acceptance rate for additions while avoiding nonphysical overlaps with other molecules that would arise from more straightforward random additions. A value of $\mu$=100 kcal/mol was employed for Zn$^{2+}$ additions and of $\mu$=500 kcal/mol for ligand additions. The latter one needs to be larger, since the bigger volume of the ligand lowers the probability of addition acceptance.
During this period, the position of all the Zn$^{2+}$ and ligand species present in the system remains fixed, while the solvent molecules are allowed to relax.
It is important to mention that the method should not be considered as an equilibration with a reservoir at chemical potential $\mu$, as occurs in Grand Canonical Monte Carlo algorithms. Instead it should be seen as way of adding a specific amount of molecules avoiding high energy configurations.
This protocol was implemented inside a unique LAMMPS input script, which is available at \url{https://github.com/rosemino/MAGNIFY/tree/main/ZIF-8_Growth_CmuMD}.

\section*{Presence of oligomeric species during the simulations}\label{oligomer}

\begin{figure}[H]
   \centering
    \includegraphics[width=0.5\textwidth]{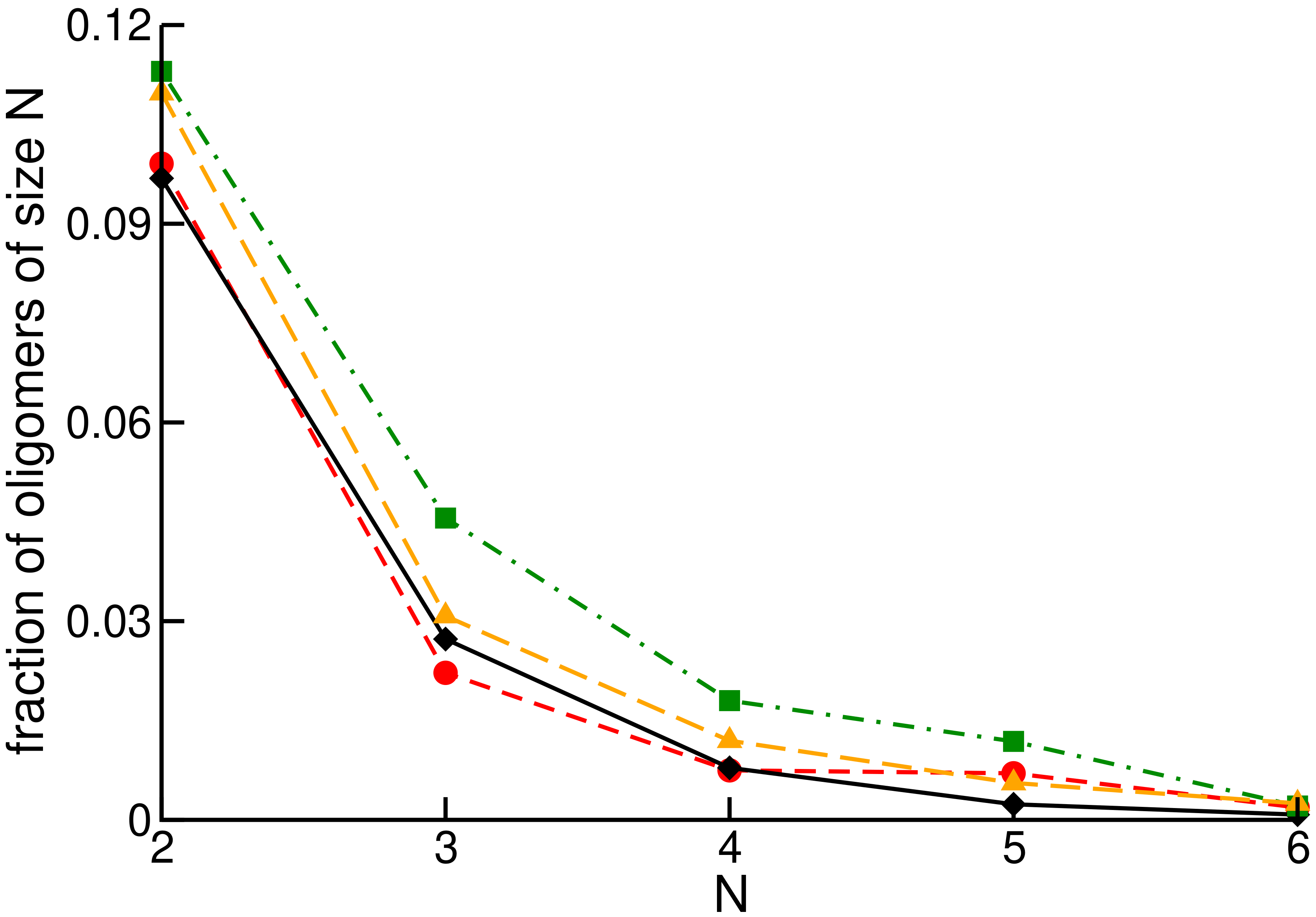}
    \vspace{0.1cm}
    \caption{Fraction of oligomers of size N>1 formed during the full simulation time in each of the conditions studied.
    Color code: [Zn$^{2+}$] = 0.1 M, T = 298 K (red), [Zn$^{2+}$] = 0.2 M, T = 298 K (black), [Zn$^{2+}$] = 0.4 M, T = 298K (green) [Zn$^{2+}$] = 0.2 M, T = 320 K (orange).}
    \label{fig:oligos}
\end{figure}

In Fig.~\ref{fig:oligos} we show the fraction of oligomers of different sizes formed in the TR at each of the thermodynamic conditions studied.
The oligomer size N includes both Zn$^{2+}$ and ligand moieties. The plot starts from N=2, as N=1 are "monomers".

\section*{Concentration dependence of the Growth rate}\label{rates}
\begin{figure}[H]
   \centering
    \includegraphics[width=0.5\textwidth]{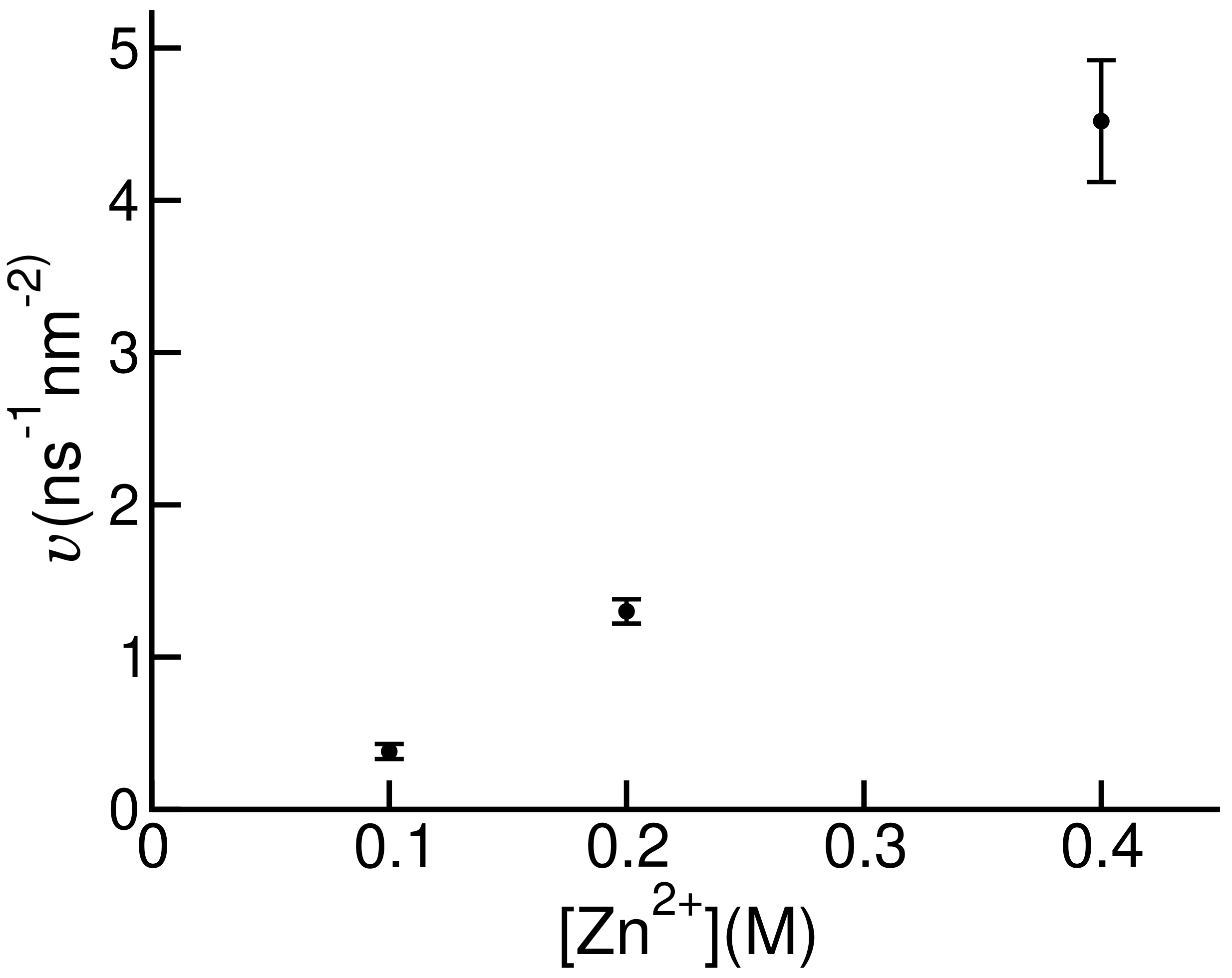}
    \vspace{0.1cm}
    \caption{Growth rates as a function of [Zn$^{2+}$] for systems at $T=$ 298 K. Error bars are computed from the dispersion of the results between independent simulations.}
    \label{fig:rates_summary}
\end{figure}

In Fig.~\ref{fig:rates_summary} we show the growth rates as a function of the ion concentration for all systems at a constant temperature of 298 K. A non linear behaviour is observed as explained in the main text. Individual growth rate values for each replica are summarized in Table~\ref{tabla2}.

\setcounter{table}{0}
\renewcommand{\thetable}{S\arabic{table}}

\begin{table}[hbtp]
\small
  \caption{Growth rates for Zn$^{2+}$ and ligands per unit of surface area for each independent simulation at each of the studied thermodynamic conditions.}
  \label{tabla2}
  \begin{tabular*}{0.9\textwidth}{@{\extracolsep{\fill}}lllll}
    \hline
  [Zn$^{2+}$](M) & $T$(K)  & Simulation tag &  $v_{\mathrm{Zn^{2+}}}$(ns$^{-1}$nm$^{-2}$)& $v_{\mathrm{MIm^-}}$(ns$^{-1}$nm$^{-2}$) \\
    \hline
0.1  & 298  & 1  & 0.11       & 0.19   \\
0.1  & 298  & 2  & 0.13       & 0.25   \\
0.1  & 298  & 3  & 0.14       & 0.31   \\
0.2  & 298   & 1 &   0.40     &  0.83     \\
0.2  & 298   & 2 &   0.41     &  0.86     \\
0.2  & 298   & 3 &   0.48     &  0.93     \\
0.4  & 298  & 1  &    1.62   &       3.27       \\
0.4  & 298  & 2  &    1.62   &       3.23       \\
0.4  & 298  & 3 &    1.26   &       2.69      \\
    \hline
  \end{tabular*}
\end{table}
\newpage

\bibliography{main}

\end{document}